\documentclass[a4paper]{spie}  %>>> use this instead for A4 paper
\usepackage[]{graphicx}

\title{Fringe Tracker for the VLTI Spectro-Imager}

\author{Leonardo Corcione\supit{a}, Donata Bonino\supit{a}, David F. Busher\supit{b}, Mario
Gai\supit{a}, Sebastiano Ligori\supit{a}, Davide Loreggia\supit{a}, Giuseppe Masssone\supit{a} and John S. Young\supit{b} \skiplinehalf
\supit{a}Istituto Nazionale di Astrofisica -- Osservatorio Astronomico di Torino, Strada Osservatorio, 20, 10025 Pino Torinese(TO) , Italy; \\
\supit{b}Cavendish Laboratory -- University of Cambridge, Cambridge, United Kingdom }

\authorinfo{Send correspondence to Leonardo Corcione: E-mail: corcione@oato.inaf.it, Telephone: +39 011 8101966}
\begin{document}
  \maketitle

%%%%%%%%%%%%%%%%%%%%%%%%%%%%%%%%%%%%%%%%%%%%%%%%%%%%%%%%%%%%%
% citations should be removed from abstract
\begin{abstract}
The implementation of the simultaneous combination of several telescopes (from four to eight) available at Very Large Telescope Interferometer (VLTI)
will allow the new generation interferometric instrumentation to achieve interferometric image synthesis  with unprecedented resolution and efficiency.
The VLTI Spectro Imager (VSI) is the proposed second-generation near-infrared multi-beam instrument for the Very Large Telescope Interferometer,
featuring three band operations (J, H and K), high angular resolutions (down to 1.1 milliarcsecond) and high spectral resolutions. VSI will be equipped
with its own internal Fringe Tracker (FT), which will measure and compensate the atmospheric perturbations to the relative beam phase, and in turn will
provide  stable and prolonged observing conditions down to the magnitude K=13 for the scientific combiner. In its baseline configuration, VSI FT is
designed to implement, from the very start, the minimum redundancy combination in a nearest neighbor scheme of six telescopes over six baselines, thus
offering better options for rejection of large intensity or phase fluctuations over each beam, due to the symmetric set-up. The planar geometry solution
of the FT beam combiner is devised to be easily scalable either to four or eight telescopes, in accordance to the three phase development considered for
VSI. The proposed design, based on minimum redundancy combination and bulk optics solution, is described in terms of opto-mechanical concept,
performance and key operational aspects.
\end{abstract}

%>>>> Include a list of keywords after the abstract


%%%%%%%%%%%%%%%%%%%%%%%%%%%%%%%%%%%%%%%%%%%%%%%%%%%%%%%%%%%%%
\section{INTRODUCTION} \label{sec:intro}
An international consortium has proposed to ESO the development of the VLTI Spectro Imager (VSI) (\cite{Malbet}), in the framework of the call for
Phase-A study of second-generation VLTI instruments. VSI has been designed to provide spectrally-resolved near-infrared images at angular resolutions
down to 1.1 milliarcsecond and spectral resolutions up to $R = 12000$, in the near infrared spectral range (J, H and K), on targets as faint as $K =
13$. % without requiring a brighter reference object in the nearby of the scientific source.
These high level performances are for achieving unprecedent milliarcsecond-resolution images of a wide range of targets with an huge potential for
breakthroughs in several astrophysics fields, e.g. probing the initial conditions for planet formation in the AU-scale environments of young stars,
mapping the chemical and physical environments of evolved stars, stellar remnants and stellar winds, disentangling the central regions of active
galactic nuclei and supermassive black holes, or imaging phenomena like convective cells on the surfaces of stars.
\par
\begin{figure}[h]
\begin{center}
\begin{tabular}{c}
  % Requires \usepackage{graphicx}
  \includegraphics[scale=0.4]{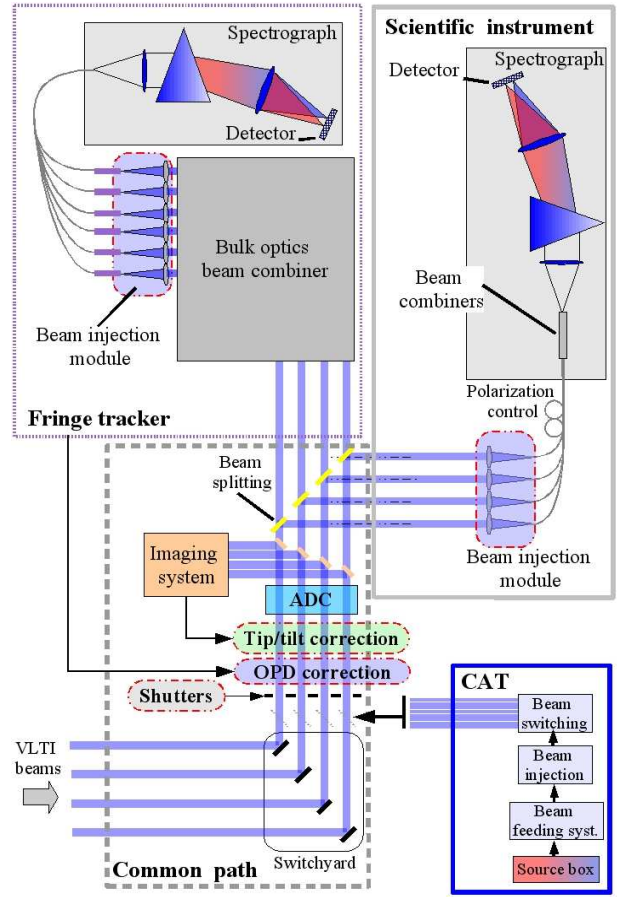}
\end{tabular}
\end{center}
  \caption[]{\label{FT-Blocks}VSI system conceptual diagram.}
\end{figure}

\noindent
VSI is designed to feature 4 main components (figure \ref{FT-Blocks}): the Science Instrument (SI), the Fringe Tracker (FT), the Common Path (CP) and
the Calibration and Alignment Tools (CAT). Science Instrument (\cite{Jocou}) features beam combination using single mode fibers, an Integrated Optics
(IO) chip and 4 spectral resolutions through a cooled spectrograph. The Common Path includes low-order adaptive optics aimed at only tip-tilt
corrections.
%The servoloop systems relax the constraints on
%the VLTI interfaces by allowing to servo optical path length differences and optimize the fiber
%injection of the input beams to the required level.
An internal optical switchyard allows to choose the best configuration of the VLTI co-phasing scheme in order to perform phase bootstrapping for the
longest baseline on over-resolved objects.

\par\noindent
The VSI internal fringe Tracker (FT) performs measurements and corrections of the atmospheric perturbations to the relative beam phase, and locks
fringes on the maximum contrast position, thus providing stable and prolonged observing conditions down to the magnitude K=13 for the scientific
combiner.
\par\noindent
In section \ref{sec:FT-Intro}, the goals of the Fringe Tracker are described, together with the operational modes that are foreseen in order to reach
them. After introducing the general concept underlying the FT, opto-mechanical concepts and key aspects of the design are presented in more details. In
section \ref{sec:SysPerf}, expected performances, optical and on-sky sensitivity, are given.

%%%%%%%%%%%%%%%%%%%%%%%%%%%%%%%%%%%%%%%%%%%%%%%%%%%%%%%%%%%%%
\section{VSI Fringe Tracker - Overview}
\label{sec:FT-Intro}
VSI internal Fringe Tracker (FT) is required to co-phase at least four telescopes, with a possible extension to six and eight
telescope arrays. VSI FT must ensure the optical path difference (OPD) compensation against atmospheric turbulence, relative to the combined beams in
co-phasing and the coherencing regimes:
\begin{description}
\item[cophasing mode:] the OPD error,associated to fringe phase fluctuations, is measured
and fed back, in real time and at high speed, either to the OPD compensator internal to VSI or to the VLTI Delay Lines, in order to stabilize baselines
down to a few tenths of working wavelength (e.g.  $\lambda/20$), thus allowing on-chip integration on the scientific instrument over several atmospheric
coherence times.
\item[coherencing mode:] the group delay (GD) error, i.e. the white light fringe
position variation, is compensated in real time at lower speed, to keep fringes stable within a fraction of the coherence length (e.g. maintaining
fringes in the coherence envelope with $\lambda$ accuracy) and the fringe contrast to adequate levels to integrate fringes incoherently over minutes.
\end{description}
The fringe contrast and the associated Signal to Noise Ratio (SNR) on fringe phase are used to evaluate the relative optical path differences between
beams, which are sent to the optical path control system in order to stabilize the internal optical paths. The FT is fed with photons from the same
scientific source under analysis by the scientific combiner of VSI: the FT measures fringe parameters by H or K photons not in use to scientific
instrument, and makes use of dispersed fringes over at least three sub-bands, for a proper implementation of both phase and group delay measurements.
\par\noindent
The FT beam combiner realizes the Minimum Redundancy Combination concept(\ref{subsec:MRC}) so as to allow the usage of N baselines instead of the
minimum N-1, and thus to link in a ring all the telescopes feeding the interferometer. The combined signals are spatially filtered by injection into
optical fibres, which, in turn, transport the signal into a cryogenic dewar. Inside the dewar, a cryogenic spectrograph allows fringes to be measured
simultaneously in a suitable number of adjacent spectral channels covering one complete spectral band, either H or K. The spectrograph light falls on
the detector, the output of which is digitized and fed to the OPD servo, which executes the FT algorithms, computes the OPD error on all the measured
baselines and feeds error signals to the internal fast OPD corrector and the VLTI delay lines.
\par\noindent
In subsections below, the proposed design, based on minimum redundancy combination and bulk optics solution, is described in more details concerning the
opto-mechanical concept, its functional performances and critical design aspects.

\begin{figure}[h]
\begin{center}
\begin{tabular}{c}
  \includegraphics[scale=0.6]{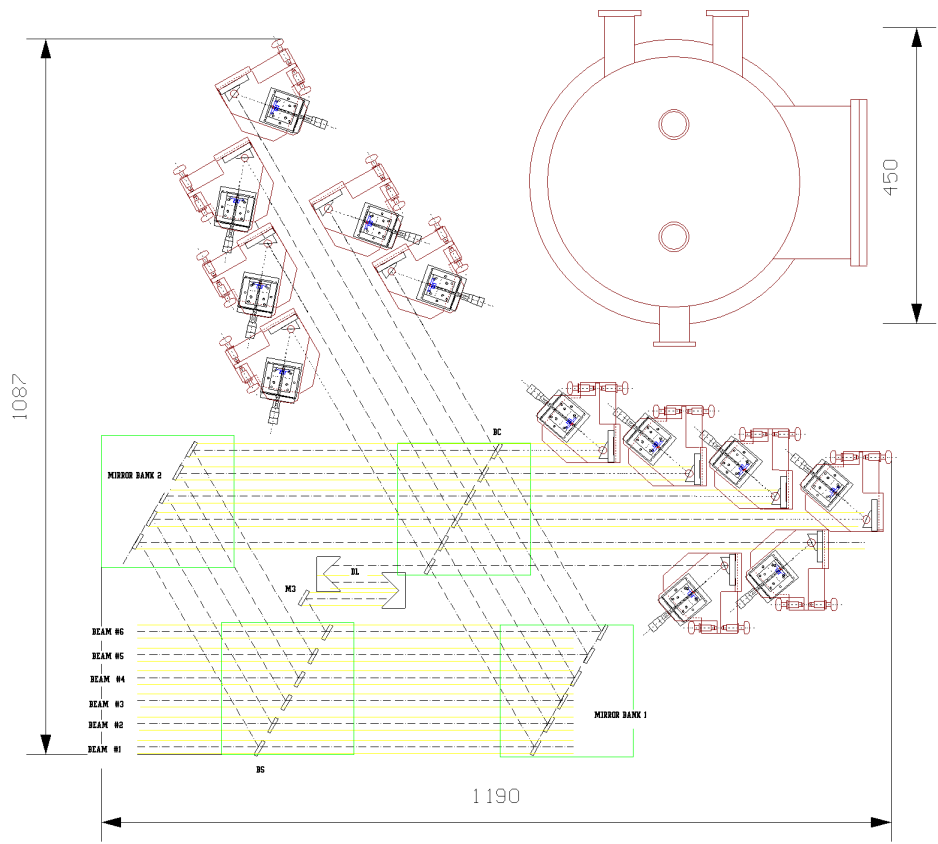}
\end{tabular}
\end{center}
\caption{\label{FT-Layout}VSI Fringe Tracker layout and footprint: six-beam combiner, fiber injection modules and cryostat}
\end{figure}

%%-----------------------------------------------------------
\subsection{Minimum Redundancy Combination} \label{subsec:MRC}
The Minimum Redundancy Combination (MRC) uses pair-wise combination instead of an all-with-all combination (\cite{Ribak}). MRC combines each beam  with
the two neighbor beams, in order to provide as many complementary outputs as the number of beams in use, and to allow simultaneous phase estimates over
the equivalent number of baselines. This is actually a redundant measurement, but the redundancy is minimal, as only one additional measurement is
performed with respect to the minimum $N-1$ value. This approach improves the robustness of the control system implementation at system level, in
particular with respect to infrastructure and environment disturbances (e.g. local seeing in on-air delay lines or mechanical vibrations). Any
occasional flux dropouts or large piston fluctuations on any of the beam can be in principle identified: the disturbance can be possibly factored out by
appropriate algebraic composition of the error measurements over the available baselines, or the best possible operation of the interferometer,
compatible with the observing condition, can be retained through the unaffected baselines. Moreover, because of the intrinsic symmetry among the
combination pairs, all the measurements for all baselines are potentially comparable in terms of SNR.
\par\noindent
The optical configuration of the FT beam combiner, realizing MRC concept for six beams (i.e adopted baseline configuration for the VSI FT), results in
planar arrangement of a suite (see Figure \ref{FT-Layout}) of beam splitter (BS), beam combiner(BC) and plane reflecting mirrors, by which each entering
beam is split into two parts and each half is combined with the light from the nearest-neighbor beam to form fringes. The minimum redundancy, i.e. the
telescope ring closure, is obtained by superposing beams 1 and 6. The planar layout, beam angles, as well as the common beam splitting and beam
combination arrangement, are preserved; beam 6 is folded by mirror M3 towards the position of beam 1 into the beam combiner, and the balance of the
optical path between the two beams is achieved by introducing an internal optical delay system.
\par\noindent
Alternative implementations of the MRC concept are provided by Ligori et al., this conference (\cite{Ligori}).

%%-----------------------------------------------------------
\subsection{Opto-mechanical layout}
\label{subsec:opto-mech} The FT beam combiner produces pairs of complementary outputs (180 degrees out of phase) and then it intrinsically implements
the static "AC" measurement scheme. The planar geometry  is arranged to allow low angles of incidence, order of 30 degrees, onto the beam splitter/beam
combiner. This allows high-efficiency dielectric beamsplitter coatings to be used, and also means that the amplitude and phase differences between the S
and P polarisations are small (typically less than 1\% in amplitude). Thus unpolarized fringes can be formed with little loss in fringe contrast.
Extension to the "ABCD" phase measurement approach, i.e. by four $\lambda/4$ spaced fringe samples, is achievable through temporal modulation of the
fringe phase. To do this, the mirrors in Mirror Bank 1 are mounted on piezoelectric actuators and their positions are stepped by a quarter of a
wavelength (at the central wavelength of the working band) in the middle of every exposure: by synchronizing the detector readout with the stepping of
the piezos, AC and BD samples from both combiner outputs can be routinely acquired. Stepping the piezos rather than scanning them means that the fringes
are not "smeared" during the detector integration time: this can cause a 10\% loss in visibility when using a linear scan instead of a stepped scan. The
size of this optical layout is a strong function of the beam-to-beam lateral spacing. The size shown in the diagram (Figure \ref{FT-Layout}) is relative
to "beam pitch" of 35$mm$; the whole footprint has been minimized assuming monolithic mounts for beamplitter, beamcombiner and mirrors.
\par\noindent
The chief advantages of this optical scheme are:
\begin{enumerate}
\item{The telescopes are closed in a ring and all baselines are measured with minimum dilution of light}
\item{Minimum number of surfaces to introduce photon losses or wavefront distortion: each beam
sees a total of two beamsplitters plus one reflection between the FT entrance and the dewar coupling optics}
\item{High-efficiency beamsplitter coatings can be used because of the low
angles of incidence}
\item{The alignment of the system is straightforward, being a simple pairwise
scheme}
\item{The system can be easily scaled for any number of beams}
\end{enumerate}

\subsection{Spatial filtering} \label{subsec:SpatialFilter}
After the combination, the beams are focused onto mono mode fibers, which carry the interferometric spots directly into the cryostat. Although not
strictly required after combination, the usage of single mode fibers is justified by several requirements. They provide high flexibility in the
arrangement of beams into the cryostat. Then they act as a spatial filter, thanks to the matching of the central lobe of the Point Spread Function (PSF)
of each beam coming from the beam combiner with the associated fiber core. Finally, previous experiences, such as Phase Referenced Imaging and
Microarsecond Astrometry (PRIMA) FSU, showed that the use of cold fibres and optics in the cryostat can reduce the thermal background noise of at least
one order of magnitude. Measurements in two bands require in principle two sets of fibre bundles, each tailored for operating bands (H or K). However,
the use of a single set of mono-mode fibers for both operational bands appears also feasible and compatible with required performances. At system design
level, the approach greatly relaxes the complexity of the signal routing towards the dewar, simplifies the configuration of relay optic for fiber
injection, and also easies the implementation of working spectral band selection, which can be accomplished at software level (subsection
\ref{subsec:ColdOptics} and \ref{subsec:SignalDet}). Either sharing a single set of K mono-mode fibre with H band or employing single mode H fiber for K
band is feasible. The former is expected to introduce fluctuations of the H beam, because of the possible coupling of higher modes to the fundamental
one; but the effect can be partially compensated or controlled at the level of camera spectrograph inside the cryostat by suited motorized cryogenic
actuator for the accurate positioning of the fibers in front of the detector. The latter, being not properly optimized for the K PSF extension,
introduces a marginal loss of coherent photons for the measurement in K band; but the equivalent reduction of the environmental background should
eventually preserve the instrument performances.

\subsection{Cold optics} \label{subsec:ColdOptics}
\begin{figure}[h]
\begin{center}
\begin{tabular}{c}
 \includegraphics[scale=0.4, angle=-90]{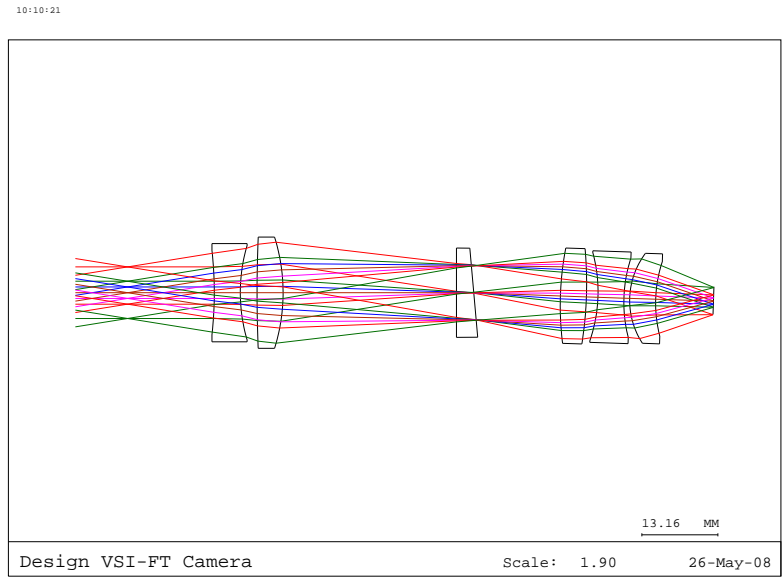}
\end{tabular}
 \end{center}
 \caption{\label{FT-ColdOptics}FT cold optics layout}
\end{figure}

Inside the dewar, fibres are properly aligned in front of a dispersive optics, which spectrally disperses spots over few, 3 to 5, pixels. The FT
spectrograph preliminary layout consists of 5 spherical lenses plus a prims (Figure \ref{FT-ColdOptics}) built with very common IR glasses (Silica and
ZnSe). The magnification is 0.6 and the system is considered telecentric. The fibres feeding the spectrograph are assumed to have a NA = 0.16 and the
distance fiber-Lens1 is equal to $15mm$.

\par\noindent
The evaluated values of the resulting spot diagram root mean square (rms) radius for each wavelength is order of 7-9 microns, in agreement with the need
to have the PSF mapped over the $18.5 \mu$m pixel size of HAWAII-1 detector, i.e. the selected detector for the VSI FT.
\par
\begin{figure}[h]
\begin{center}
\begin{tabular}{c|c}
 \includegraphics[scale=0.3]{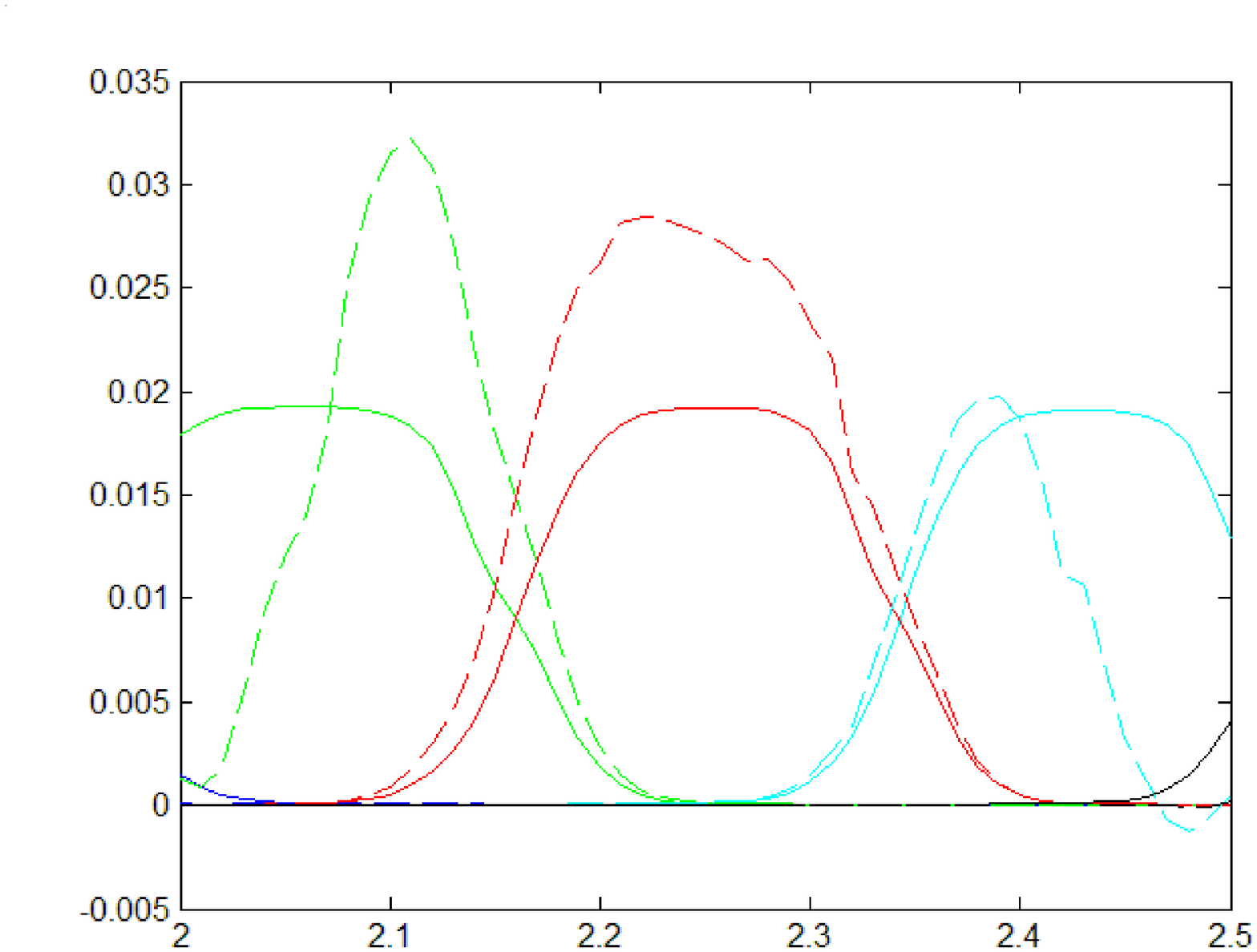} & \includegraphics[scale=0.3]{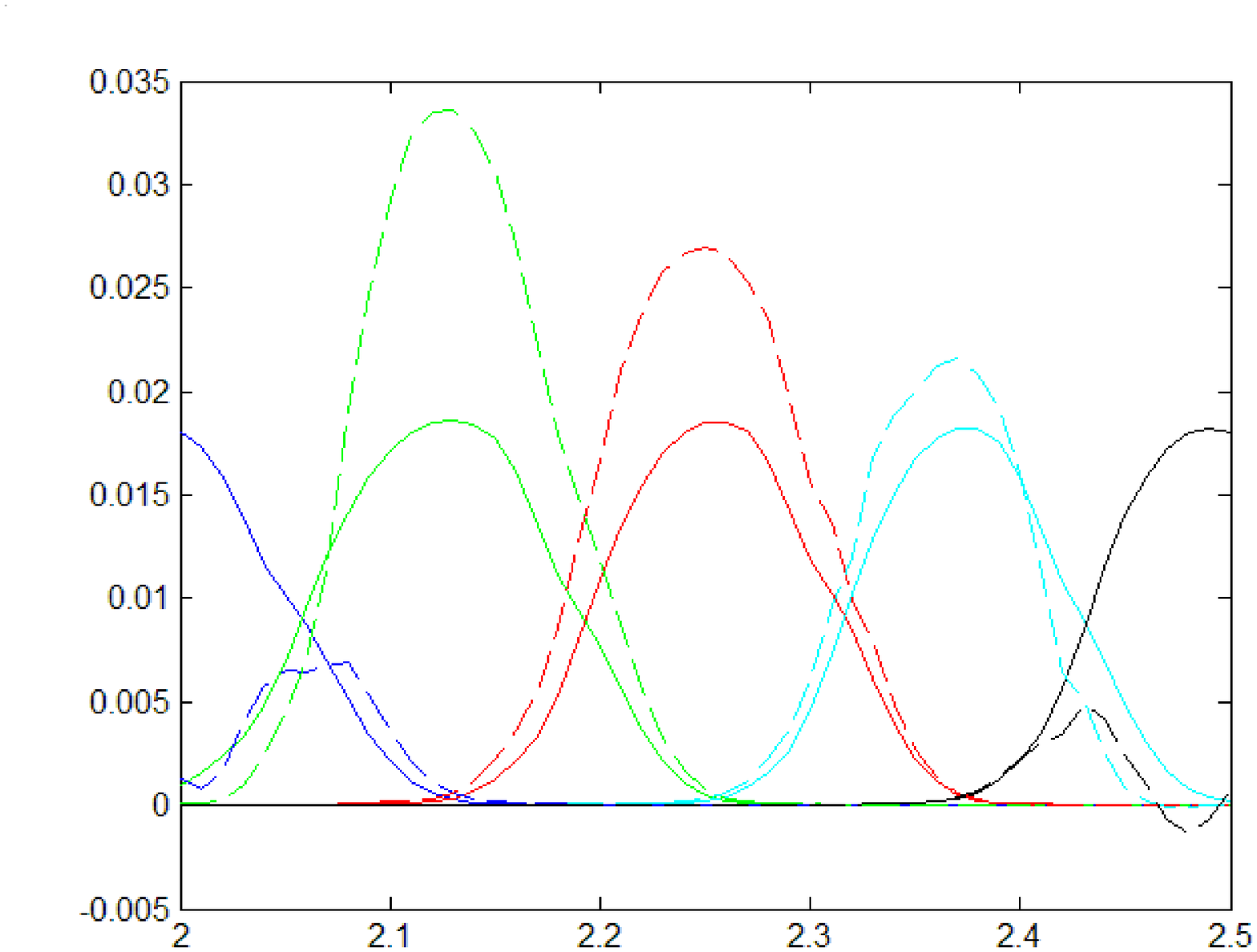}
 \end{tabular}
\end{center}
  \caption[]{\label{FT-3vs5SpectralDispersion} FT spectral response: 3 band case (left); 5 band case(right).
  The intrinsic optical dispersion is shown in solid lines; dispersions in dashed include transmission and source spectrum}
\end{figure}

\noindent
The order of spectral dispersion needs careful assessment to make the tracking algorithm sensitive and reliable: the higher the resolution is,
potentially the more sensitive the tracking algorithm is to large fringe jump; but with less photons in every spectral element and the fringe
measurements are likely accomplished at reduced SNR. Still, the tracking control loop bandwidth also depends on the number of pixels to read (subsection
\ref{subsec:SignalDet}). In the framework of PRIMA FSU like instrument configuration, provisional analysis -also based on reported atmospheric and optic
transmissions, as well as on quoted detector quantum efficiency- shows how splitting the working band, either H or K, in five sub-bands makes available
just three central wavelengths for efficient usage in the measurements (Figure \ref{FT-3vs5SpectralDispersion}: effective wavelengths at 2.02, 2.16 and
2.32$\mu$m for the K band case, with spectral resolution 35, 26 and 32 respectively). The signal is actually concentrated mostly in the three central
sub-bands, and the nearby pixels receive a marginal fraction of the signal, which shall be further minimized during design optimization.
\par\noindent
The usage of three wavelengths for each band is the adopted baseline solution  for the VSI FT operations, it allows a good trade-off between algorithm
robustness and control loop bandwidth: adequate photon distribution among three bands allows measurements at high SNR for accomplishing cophasing (from
photons falling in the central band) and coherencing (from photons from lateral bands); reduced number of pixels ensures fast readout speed and high
control bandwidth.
\par\noindent
Assuming to arrange the spectral dispersion such that the whole H+K band of each combiner output is continuously distributed over 6 pixels, i.e. 3 pixel
dispersion per band, a complete scan for the six-telescope configuration requires to read at the most 72 detector pixels (twice for ABCD fringe coding),
and the working wavelength selection is achievable at pixel processing level, e.g. clocking only pixels associated to the working band, with no need of
hardware reconfiguration inside and outside the dewar.

\subsection{Signal detection}\label{subsec:SignalDet}
\begin{figure}[h]
\begin{center}
\begin{tabular}{c}
 \includegraphics[scale=0.5]{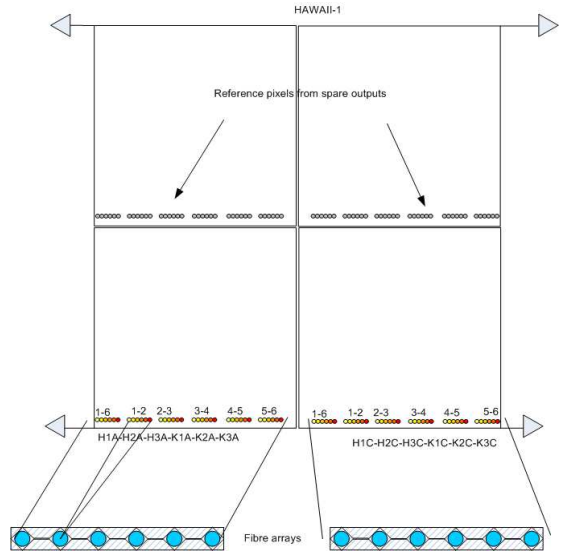}
 \end{tabular}
\end{center}
  \caption[]{\label{FT-Detection}FT detection scheme by HAWAII-1 detector.}
\end{figure}

Signals are properly mapped and processed, according to detector's pixel architecture, in order to allow measurements at high SNR and high speed.
HAWAII-1 detector from Teledyne Imaging Sensors (former Rockwell) is the sensors selected for the VSI FT; the pixel array (1024x1024 pixels,
$18.5{\mu}m$ in size) is arranged in 4 quadrants - 512x512 pixels in size, with independent outputs. The proposed linear, one dimensional configuration
(Figure \ref{FT-Detection}) of the spectrograph is mapped onto the quadrant architecture by focusing the complementary outputs -phase A and phase C
signals- of the beam combiner onto two adjacent quadrants, i.e. 6 spectra per quadrant; spectra are distributed along a common pixel row, with a
horizontal inter-spectra guard-gap, within a single quadrant, of few pixels, e.g. 3 pixel wide.
\par\noindent
Reference pixel values (for effective data processing, e.g. bias subtraction) can be either taken from some -one or two- of the inter-band pixels, with
negligible increase in readout time of one or two pixel time, or acquired in parallel from spare outputs with no increase in readout time.
\par\noindent
In the baseline configuration, with two bands and 3-pixel dispersion per band, six beams and six controlled baselines, and with the extension of the
in-hardware AC fringe sampling to the ABCD measurement scheme through temporal modulation, a complete fringe scan involves 2 detector outputs and 36
pixels per output plus some reference pixels, but only half of them need to be read, accordingly to the band selection.

\par\noindent
Tracking algorithm performance are set by sensors intrinsic features, as well as signal processing technique in use.
\par\noindent
Measurement SNR definitively depends on the adopted detector readout mode:
\begin{description}
\item [CDS Mode:] In conventional Correlated Double Sampling (CDS) mode,
the HAWAII-1 elementary readout noise (RON) is of 15 electrons rms at the maximum pixel sampling allowed rate of 500KHz.
\item[NDRO Mode:] Non-Destructive Readout(NDRO) technique is associated to multiple sampling of the signal
along the detector integration ramp and to the linear fit of the collected signal points. If applied with $N$ number of samples, the technique reduces
the effective RON by a factor of $\sqrt{\frac{6(n-1)}{n(n+1)}}$ with respect to the figure of the basic CDS scheme; with order of 100 samples, the
effective RON, potentially achievable by HAWAII-1, is lower then 4-5 electrons. Depending on the number of samples involved, NDRO approach is eventually
applicable to the FT in quite good atmospheric condition regimes (e.g. coherence time longer than 2 ms).
\end{description}
Concerning control loop bandwidth, a single readout sequence for the AC coding takes 72 micro-seconds, to read 72 pixels (H+K dispersed bands) equally
divided between two parallel outputs; CDS mode needs twice those times, i.e. 144 $\mu$s at the maximum sampling rate HAWAII-1 allows. Allocating a
convenient overhead for skipping the pixels of inter-band gap (about 1$\mu$s per gap, applicable only to the quadrant detector) and for dead time (e.g.
one frame time, order of 100$\mu$s) for stepping the internal optical path, the maximum rate sustainable for a complete set of "ABCD" readings, i.e. the
fringe tracking control loop bandwidth, is realistically close to 2KHz, which rises to 2.8KHz if the readout involves only the spectral channels
associated to the selected working wavelength, i.e 18 pixels per output instead of 36.
\par\noindent
The performance is well compatible with reasonable values of the fringe coherence time, quoted as $4 ms$ in H and $6 ms$ in K; it allows for a fringe
sampling rate (equivalent to the 2-step modulation rate of the piezo modulator of the FT beam combiner) from eight to twelve time faster the expected
atmospheric fringe jitter, ultimately set by the weather conditions.

%%%%%%%%%%%%%%%%%%%%%%%%%%%%%%%%%%%%%%%%%%%%%%%%%%%%%%%%%%%%%
\section{Instrument performance} \label{sec:SysPerf}
The potential performance of the FT are analyzed taking into account instrumental factors and the current performance of VLTI. The instrumental
contribution (i.e. system optical throughput, visibility loss, thermal background and detector noise) is evaluated considering the FT set-up described
above; the contribution from components deployed along the VSI common path are also included in the overall instrument throughput.

%%----------------------------------------------------------------
\subsection{FT optical throughput}\label{subsec:OptThr}
For the evaluation of the optical throughput budget, a percentage of flux loss is estimated for each reflective surface (2\%), each transmissive
component (1\%), each beamsplitter (2\% uncertainty on the flux division), for the detector quantum efficiency (60\%), resulting in a total internal
loss of 55\% in H band and 59\% in K band induced by optical surfaces. The other effects and the total system throughput are summarized in Table
\ref{FT-Throughput}.

\begin{table}[h]
  \centering
\begin{tabular}{lcc}
Wavelength & H & K \\
\hline
Surface and transmission losses & 0,55 & 0,59 \\
Spatial filtering coupling loss for 100\% Strehl wavefront & 0,80 & 0,80 \\
Spatial filtering coupling loss due to high-order WFE & 0,83 &  0,91 \\
Reflection losses, fibre ends & 0,9 & 0,9 \\
Coupling loss to spatial filter: tip/tilt misalignments & 0,95 & 0,95 \\
Coupling loss to spatial filter: pupil  misalignments & 0,98 & 0,98 \\
System transmission & 0,31 & 0,37 \\
Spectrograph-pixel coupling & 0,95 & 0,95 \\
Detector QE & 0,60 &  0,60 \\
FT total throughput & 0,18 &  0,21 \\
\end{tabular}
  \caption{FT optical throughput}\label{FT-Throughput}
\end{table}

%%----------------------------------------------------------------------------
\subsection{FT limiting magnitude} \label{subsec:SNR}
Hereafter, the main factors affecting the Signal to Noise Ratio (SNR) and, in turn, the on-sky instrument performance are described and quantified in
order to provide a provisional estimate of the operational limiting magnitude of the VSI FT, in context of VLTI.
\par\noindent
The SNR is computed in accordance to the formula below(\cite{LeBouquin}), assumed as common computational framework for VSI.

\begin{equation}\label{eq:SNR}
    SNR = \frac{N_{AC}^2V^2}{\sqrt{\sigma_{ph}^2 + \sigma_{det}^2}}
\end{equation}
\par\noindent
$N_{AC}$ is the number of coherent source photons effectively employed in the fringe measurements, i.e. the modulated part of the fringe signal. In a
pair-wise combination, it is formed by the correlation of the fraction $\alpha$ of the beams coming from the two given apertures; assuming all apertures
receive the same amounts of photons N, $N_{AC} = 2\alpha N$.
\par\noindent
V is the system visibility, inclusive visibility loss from VLTI infrastructure. The visibility can be degraded by several effects, both observational,
such as wavefront corrugations and piston jitter, and instrumental (e.g. intensity mismatches due to optical misalignment). From preliminary laboratory
results on a prototype combiner, the overall degradation of fringe contrast is conservatively quoted less than 5\% in broadband unpolarised light.
\par\noindent
$N_{DC}$ is the number of incoherent photons due to the source, i.e. the signal continuum which is subtracted but contribute to the noise. It is the sum
of all beams coming from contributing telescopes $n_T$, i.e $N_{DC} = n_T\alpha N$, with $n_{DC}$ the number of telescopes contributing to the DC part:
in a pair-wise combination $n_T = 2$.
\par\noindent
The noise terms $\sigma_{ph}$ and $\sigma_{det}$ are respectively the photon noise from the source and the background, and the total read-out noise:
\begin{equation}\label{eq:Noise}
    \sigma_{ph}^2 = (N_{DC}+N_{BG})^2+2(N_{DC}+N_{BG})N_{AC}^2V^2; \hspace{1cm} \sigma_{det}^2 = 2n_p^2\sigma_e^4.
\end{equation}
\par\noindent
$N_{BG}$ is the number of background photons, $\sigma_e$ is the read-out noise per pixel and $n_p$ is the number of pixel to be involved in the
measurement.
\par\noindent
With respect to the SNR expression and computational framework it refers to, the involved quantities must be computed taking into account the
instrumental parameters (subsection \ref{subsec:OptThr}) and the details of implementation of the beam combination.
\par\noindent
In pair-wise nearest-neighbor combination scheme, each beam is split in two equal intensity fractions, then each measurement channel receives half the
photons N collected by single aperture. Therefore, for the sake of the SNR computation, $n_T = 2$, independently from the total number of telescopes
combined, and  $\alpha =1/2$; then, $N_{AC}=N$ and $N_{DC}=N$ as well.
\par\noindent
The total number of pixels depends upon the details of  fringe tracking algorithm implementation: for the  ABCD coding with 3 spectral channels, the
usage of the whole spectral band imposes to read three times the number of pixels involved into signal phase coding; so, the effective number of pixels
to read is 12.
\par\noindent
The adopted RON figure is the one quoted for the HAWAII detector operated in CDS readout mode, i.e 15 electrons rms per read.
\par\noindent
The thermal background coming from the combination of the telescope, the sky background and all the warm optics in the beam train can be modeled as a
single blackbody at approximately 300° K, emitting $3.64\star 10^{15} photons/second/sr/m^2$  over the K band (2.04-2.37 $\mu$m). In the spatial
filtering regime by a fibre diameter D which is exactly matched to the first Airy ring of the diffraction limited spot corresponding to that pupil, the
beam \'{e}tendue is given by
\begin{equation}\label{eq:etendue}
\pi(1.22\lambda/D)^2\pi(D/2)^2 \simeq 3.67\lambda^2
\end{equation}
In K band, effective wavelength 2.2$\mu$m, this translates to $6.9\star10^4$ photons/s in the system per beam.
\par\noindent
Photon flux at the entrance of the fringe tracker is computed starting from top atmosphere zero magnitude flux collected by the UT pupil, and assuming
the reported VLTI performance: the VLTI relevant figures used in the evaluation are summarized in Table \ref{VLTI-Performance} (taken from ESO VTL
Interface Control Document).

\begin{table}[h]
  \centering
\begin{tabular}{lcc}
Wavelength & H & K \\
\hline
Zero-mag top atmosphere flux(UT case)[ph/s] & 1.52 10(11) & 8.4 10(10) \\
Atmospheric transmission & 0.97 & 0.90 \\
VLTI Strehl Ratio (MACAO, V=13) & 0.39 & 0.47 \\
Strehl Ratio loss(21 mas tip/tilt residual) & 0.68 & 0.81 \\
VLTI Throughput &  0.11 &  0.12 \\
VLTI Visibility(0.5 ms piston jitter) & 0.9 & 0.95 \\
VLTI Visibility(5 ms piston jitter) & 0.8 & 0.9 \\
VLTI Visibility(10 ms piston jitter)& 0.7 & 0.8 \\
\end{tabular}
  \caption{FT optical throughput}\label{VLTI-Performance}
\end{table}
\par
\begin{figure}[h]
\begin{center}
\begin{tabular}{c}
  \includegraphics[scale=0.6]{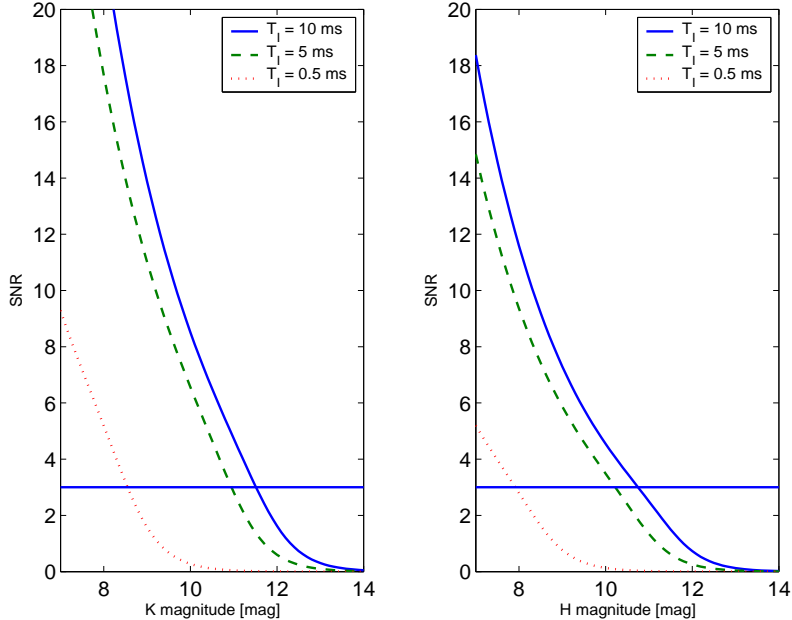}
 \end{tabular}
\end{center}
  \caption{\label{FT-SNR}VSI Fringe Tracker SNR performance as a function of magnitude and integration time: horizontal line corresponds to SNR = 3}
\end{figure}

\noindent
The resulting FT performance at VLTI with Unit Telescopes (8 meter diameter) are presented in Figure \ref{FT-SNR}, where the achievable SNR in H and K
band is shown as a function of the magnitude. The Table \ref{FT-LimitingMag} summarizes the expected FT limiting figures in working bands and in
operating regimes applicable to cophasing ($T_{exp}\leq 1 ms$) or coherencing ($T_{exp}\geq 5 ms $), at adequate SNR levels.

\begin{table}[h]
  \centering
\begin{tabular}{|c|c|c|c|c|}
  \hline
  % after \\: \hline or \cline{col1-col2} \cline{col3-col4} ...
  & \multicolumn{2}{|c|}{H} & \multicolumn{2}{|c|}{K} \\
  \hline
  $T_{exp}$[ms] & SNR=5 & SNR=3 & SNR=5 & SNR=3 \\
  \hline
  0.5 & 6.9 & 7.9 & 8.0 & 8.5 \\
  1.0 & 7.6 & 8.5 & 8.7 & 9.2 \\
  5.0 & 9.3 & 10.2& 10.4& 10.9 \\
  10.0& 9.8 & 10.8& 10.9& 11.5 \\
  \hline
\end{tabular}
  \caption{FT Limiting magnitude in H and K}\label{FT-LimitingMag}
\end{table}
\par\noindent
Coherent cophasing  at 1 KHz rate is performable  up the 8 mag in H and up to 9 in K; coherent coherencing, i.e. group delay tracking regime at 200Hz or
lower rate(3rd and 4th row of table \ref{FT-LimitingMag} above), can occur below the 10th mag in H, or on object fainter than 11th mag in K. Fringe
tracking at fainter magnitudes is potentially achievable through the incoherent coherencing, where the group delay is integrated incoherently over many
coherent integrations, performed in starved photon regime. It is demonstrated that the group delay tracking is possible down to SNR values per
integration of approximately 0.7 (\cite {Busher}). A conservative performance estimate in photon starved regime can be obtained by the framework  and
provisions above: assuming 20 ms incoherent integration and a SNR=1, associated to the single coherent integration, the FT limiting magnitudes are
roughly close to H=12.7 and K=13.2.

%%%%%%%%%%%%%%%%%%%%%%%%%%%%%%%%%%%%%%%%%%%%%%%%%%%%%%%%%%%%%%%%%%%%%%%%%%%%%
\section{Conclusions}\label{sec:Conclusion}
The VSI fringe tracker, based on the planar MRC concept and bulk optics solution, will allow VSI, and possibly also other VLTI 2nd generation
instrumentation, to benefit of prolonged stable observational conditions up to the 8 mag in H band (or 9 mag in K), with stable fringes at sub-$\lambda$
accuracy (co-phasing regime); while, the fringe stabilization in coherent mode is achievable, in acceptable atmospheric conditions, at one or two mag
fainter. By an incoherent approach, as for instance by software numerical integration over short coherent exposures, VSI FT could operate on sources as
faint as the 13th mag.
\par\noindent
The planar geometry solution of the VSI fringe tracker six-beam combiner is devised to be easily scalable either to four or eight telescopes, so as to
support the foreseen development phase of the VLTI infrastructure; it also provides the very appealing capability to extend, by few changes in FT
design, the dual-feed mode of PRIMA up to the eight-beam VLTI configuration, with two sets of four out of eight beams feeding the FT.

%%%%%%%%%%%%%%%%%%%%%%%%%%%%%%%%%%%%%%%%%%%%%%%%%%%%%%%%%%%%%
\acknowledgments     %>>>> equivalent to \section*{ACKNOWLEDGMENTS}
We acknowledge the financial support of INAF - ref. 1478 (2005) for contributions to the FT lab prototype.
% This unnumbered section is used to identify those who have aided the authors in understanding
%or accomplishing the work presented and to acknowledge sources of funding.

%%%%%%%%%%%%%%%%%%%%%%%%%%%%%%%%%%%%%%%%%%%%%%%%%%%%%%%%%%%%%
%%%%% References %%%%%

\end{document}